\documentclass[11pt,reqno]{amsart}

\usepackage{graphics,cancel,graphicx}
\usepackage{epsfig,psfrag,manfnt}
\usepackage{bbm,subfigure,rotating,bm,float,mathdots,wasysym}

\usepackage[all,cmtip]{xy}
\usepackage{amssymb,amsmath,amsfonts,amsthm,color}

\newtheorem{theorem}{Theorem}[section]

\theoremstyle{definition}

\theoremstyle{definition}

\theoremstyle{definition}

\begin{document}



\title[Relativistic velocity addition]{Alternative proofs for\\  
Kocik's Geometric Diagram \\
for Relativistic Velocity Addition}

\author[A. Sasane]{Amol Sasane}
\address{Department of Mathematics \\London School of Economics\\
    Houghton Street\\ London WC2A 2AE\\ United Kingdom}
\email{sasane@lse.ac.uk}

\author[V. Ufnarovski]{Victor Ufnarovski}
\address{Center for Mathematical Sciences \\Lund University\\
    221 00 Lund\\ Sweden}
\email{victor.ufnarovski@math.lth.se}

\begin{abstract} 
A geometric construction for the Poincar\'e formula for relativistic addition of velocities in one dimension 
 was given by Jerzy~Kocik in {\em Geometric Diagram for Relativistic Addition of Velocities}, 
 American Journal of Physics, volume 80, page 737, 2012. While the proof given there used Cartesian 
 coordinate geometry, three alternative  approaches are given in this article: a trigonometric one, 
 one via Euclidean geometry, and one using projective geometry. 
\end{abstract}

\maketitle

\section{Introduction}

\noindent Imagine a train moving at speed $u$ with respect to the ground 
(as reckoned by someone sitting on the ground), and further that a person P is 
running with a speed $v$ on the train (as reckoned by somebody sitting in the train). Before 1905, 
 Newtonian physics dictated that the speed of the person P as observed by someone on the ground is
 $u+v$, while we now know better; the relativistic formula for velocity addition says that the  
  speed should be  
  $(u\oplus v) :=(u+v)/(1+uv), 
 $ 
  in units in which the speed of light is $1$.

 \begin{figure}[H]
   \center
   \psfrag{A}[c][c]{${  A}$}
   \psfrag{B}[c][c]{${  B}$}
   \psfrag{U}[c][c]{${ \! V}$}
   \psfrag{V}[c][c]{${  U\;}$}
   \psfrag{u}[c][c]{${  V'}$}
   \psfrag{v}[c][c]{${  U'}$}
   \psfrag{s}[c][c]{${ \!\!\!\!\!\!\!W}$}
   \psfrag{e}[c][c]{${  u}$}
   \psfrag{f}[c][c]{${  v}$}
   \psfrag{O}[c][c]{${  O}$}
   \psfrag{C}[c][c]{${  C}$}
   \includegraphics[width=6cm]{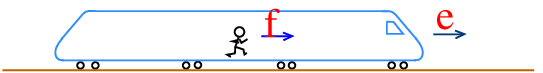}
\end{figure}

 \noindent In \cite{K}, a geometric diagram for the construction of $u\oplus v$ from $u$ and $v$ was given. We recall it below. 
 
  \begin{theorem}[\cite{K}] \label{theorem_Kocik} 
  Draw a circle with center $O$ and radius $1$. Mark points $U,V$ 
 at distances $u,v$ from $O$ 
  along the radius $OC$ perpendicular to a diameter $AB$. Let the line joining $B$ to $V$ 
  meet the circle at $V'$, and let the line joining $A$ to $U$ meet the circle at $U'$. 
  Then $u\oplus v=OW$, where $W$ the point of intersection of $U'V'$ with the radius $OC$. 
  \end{theorem}
  
 \begin{figure}[H]
   \center
   \psfrag{A}[c][c]{${  A}$}
   \psfrag{B}[c][c]{${  B}$}
   \psfrag{U}[c][c]{${ \! V}$}
   \psfrag{V}[c][c]{${  U\;}$}
   \psfrag{u}[c][c]{${  V'}$}
   \psfrag{v}[c][c]{${  U'}$}
   \psfrag{s}[c][c]{${ W}$}
   \psfrag{O}[c][c]{${  O}$}
   \psfrag{C}[c][c]{${  C}$}
   \includegraphics[width=5.4cm]{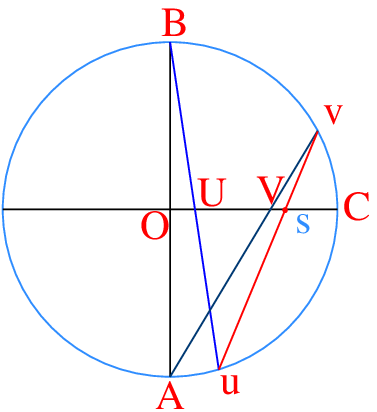}
\end{figure} 
 
\noindent This construction allows visual justification of  
the following properties of $\oplus$. For all $u,v\in [0,1]$, 
 $u\oplus v\in [0,1]$, $v\oplus 1=1$, $v\oplus 0=v$, and 
when $0\leq u, v\ll 1$, then $u\oplus v\approx u+v$. For example, let us justify this last fact geometrically. 
If $u,v\ll 1$, then $\angle OBV\approx 0$, and $AV'$ is almost parallel to $OV$.  

 \begin{figure}[H]
   \center
   \psfrag{A}[c][c]{${  A}$}
   \psfrag{B}[c][c]{${  B}$}
   \psfrag{U}[c][c]{${ \! V}$}
   \psfrag{V}[c][c]{${  U\;}$}
   \psfrag{u}[c][c]{${  V'}$}
   \psfrag{v}[c][c]{${  U'}$}
   \psfrag{s}[c][c]{${ W}$}
   \psfrag{O}[c][c]{${  O}$}
   \psfrag{C}[c][c]{${  C}$}
   \includegraphics[width=5.4cm]{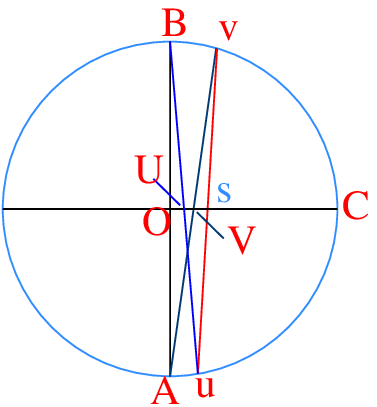}
\end{figure} 

  \noindent So $\Delta BOV$ is almost similar to $\Delta BAV'$, giving 
  $$
  AV'\approx \frac{AB}{OB}\cdot OV= \frac{2}{1} \cdot OV=2v.
  $$
   Since $AV'$ is almost parallel to $OC$,  $\Delta U'UW$ is almost similar to $\Delta U'AV'$. 
   Moreover, as $u,v\ll 1$, $U'V'\approx AB=2$, and $U'W\approx OB=1$. 
    Hence 
    $$
    UW\approx \frac{U'W}{U'V'}\cdot AV'\approx \frac{1}{2}\cdot 2v = v.
    $$
 Thus if $w:=OW$, then $w-u=UW\approx v$, that is, $w\approx u+v$.

In \cite{K}, Theorem~\ref{theorem_Kocik} was proved using Cartesian coordinate geometry.  
In the next three sections, we give three alternative proofs of this result. (The more proofs, 
the merrier!)

\section{A trigonometric proof}

\vspace{-0.18in} 

\begin{figure}[H]
   \center
   \psfrag{A}[c][c]{${  A}$}
   \psfrag{B}[c][c]{${  B}$}
   \psfrag{U}[c][c]{${ \! V}$}
   \psfrag{V}[c][c]{${  U\;}$}
   \psfrag{u}[c][c]{${  V'}$}
   \psfrag{v}[c][c]{${  U'}$}
   \psfrag{W}[c][c]{${ W}$}
   \psfrag{e}[c][c]{${  u}$}
   \psfrag{f}[c][c]{${  v}$}
   \psfrag{a}[c][c]{${  \alpha}$}
   \psfrag{b}[c][c]{${   \beta}$}
   \psfrag{O}[c][c]{${  O}$} 
   \psfrag{C}[c][c]{${  \!\!C}$}
   \includegraphics[width=5.4cm]{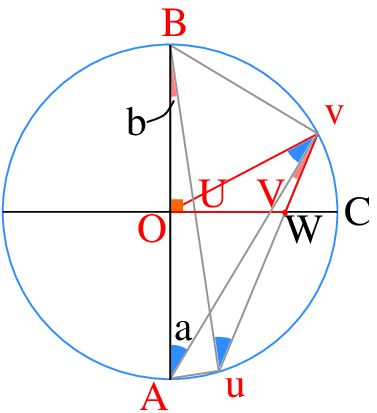}
\end{figure} 

\vspace{-0.125in} 

\noindent We refer to the picture above, calling 
$$
\angle BAU'=\angle OAU=:\alpha\quad\textrm{and}\quad \angle ABV'= 
\angle OBV=:\beta.
$$
Let  $W$ be the point of intersection of $U'V'$ and $OC$, and set $OW=:w$.  Then 
by looking at the right triangles $\Delta BOV$ and $\Delta AOU$, we see that 
$$
\tan \beta=v\textrm{ and }\tan \alpha=u.
$$
Using the Sine Rule in $\Delta OWU'$, we have 
$$
\frac{1}{\sin \angle OWU'}=\frac{OU'}{\sin \angle OWU'}
=
\frac{OW}{\sin \angle OU'W}=\frac{w}{\sin \angle OU'W},
$$
giving 
\begin{equation}\label{eq_trig_prf}
w=\frac{\sin \angle OU'W}{\sin \angle OWU' }.
\end{equation}
The proof will be completed by showing (below) that $\angle OU'W=\alpha+\beta$ and $\angle OWU'
=90^\circ+(\alpha-\beta)$, so that  \eqref{eq_trig_prf} yields 
\begin{eqnarray*}
w&=&\frac{\sin (\alpha+\beta)}{\sin(90^\circ+(\alpha-\beta))} 
=\frac{\sin (\alpha+\beta)}{\cos (\alpha-\beta)} 
=
\frac{(\sin \alpha)(\cos \beta)+(\cos \alpha)(\sin \beta)}{
(\cos \alpha)(\cos \beta)+(\sin \alpha)(\sin \beta)}
\\
&=&
\frac{\tan \alpha +\tan \beta}{1+(\tan\alpha)(\tan \beta)}
=
\frac{u+v}{1+uv},
\end{eqnarray*}
as desired. 

 First we will show $\angle OU'W=\alpha+\beta$. Note that 
 $\Delta OAU'$ is isosceles with $OA=OU'=1$ and so   
 $\angle OU'U=\angle OAU=\alpha$. The chord $AV'$ subtends equal angles at $B$ and $U'$, 
 and so $\angle UU'W=\angle ABV=\beta$. Hence 
 $$
 \angle OU'W=\angle OU'U+\angle UU'W=\alpha+\beta.
 $$
 Next, let us show that $\angle OWU'
 =90^\circ+(\alpha-\beta)$. 
 To this end, note that $\angle OUU'$ is the common exterior angle for $\Delta AOU$ and $\Delta OU'U$, 
 and using the fact that this equals the sum of the opposite interior angles in each triangle, 
 we obtain 
 $$
 90^\circ +\alpha=\angle OUU'=\beta+\angle UWU',
 $$
 so that $\angle OWU'=\angle UWU'=90^\circ+(\alpha-\beta)$, completing the proof. 

Yet another trigonometric proof can be obtained by focussing on $\Delta U'CW$, 
determining all its angles, and the side length $U'C$ (using the isosceles triangle $\Delta OU'C$), 
 enabling the determination of $WC$ ($=1-w$). The details are as follows. 
 In the isosceles triangle $\Delta OU'C$, we have 
 $$
 \angle U'OC=90^\circ-\angle BOU'=90^\circ-2\angle BAU'
 =90^\circ-2\alpha.
 $$
 As $OU'=OC=1$, we obtain $\angle OCU'=45^\circ+\alpha$ and  $U'C=2\cos (45^\circ+\alpha)$. 
 Also $\angle WU'C=\angle V'U'C=\angle V'BC= \angle ABC-\angle ABV'=45^\circ-\beta$. 
 This yields $ \angle U'WC=180^\circ- (\angle WU'C+\angle U'CW)=90^\circ+(\beta-\alpha)$. 
 Again, by the Sine Rule, this time in $\Delta U'WC$, we have 
 $$
 \frac{1-w}{\sin \angle WU'C}=\frac{WC}{\sin (45^\circ-\beta)}=\frac{U'C}{\sin \angle U'WC} =
 \frac{2\cos (45^\circ+\alpha)}{\sin (90^\circ +(\beta-\alpha))},
 $$
 that is, 
 \begin{eqnarray*} 
  1-w&=& \frac{2\cos (45^\circ+\alpha)\sin (45^\circ-\beta)}{\sin (90^\circ +(\beta-\alpha))}= 
  \frac{(\cos \alpha -\sin \alpha)(\cos \beta-\sin \beta)}{(\cos \beta)( \cos \alpha)+
  \sin \alpha )(\sin \beta)}\\
  &=&\frac{(1-\tan \alpha)(1-\tan\beta)}{1+(\tan \alpha)(\tan \beta)}
  =\frac{(1-u)(1-v)}{1+uv},
 \end{eqnarray*}
 which, upon solving for $w$, gives $w=\displaystyle\frac{u+v}{1+uv}$.

\section{A Euclidean geometric proof}
\label{section_2}
 
 \vspace{-0.18in} 
 
\begin{figure}[H]
   \center
   \psfrag{A}[c][c]{${  A}$}
   \psfrag{B}[c][c]{${  B}$}
   \psfrag{U}[c][c]{${ \! V}$}
   \psfrag{V}[c][c]{${  U\;}$}
   \psfrag{u}[c][c]{${  V'}$}
   \psfrag{v}[c][c]{${  U'}$}
   \psfrag{W}[c][c]{${  W}$}
   \psfrag{e}[c][c]{${  u}$}
   \psfrag{f}[c][c]{${  v}$}
   \psfrag{O}[c][c]{${  O}$}
   \psfrag{o}[c][c]{${  O'\;}$}
   \psfrag{C}[c][c]{${  C}$}
   \includegraphics[width=5.4cm]{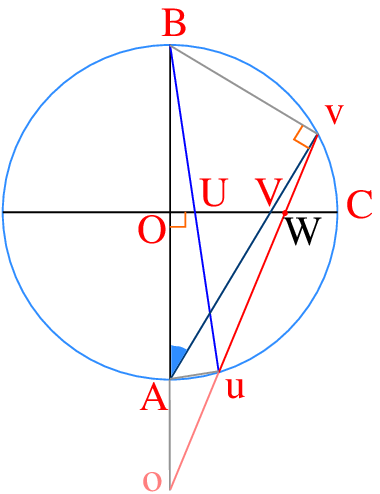}
\end{figure} 

\vspace{-0.125in} 

\noindent 
  As $\angle AOU=90^\circ=\angle AU'B$ and $\angle OAU=\angle U'AB$ (common), by the AA Similarity Rule, 
$\Delta AOU\sim \Delta AU'B$. So 
$$
\frac{AU'}{2}=\frac{AU'}{AB}=\frac{AO}{AU}=\frac{1}{\sqrt{1+u^2}},
$$
giving $AU'=2/\sqrt{1+u^2}$. Hence 
$$
UU'=AU'-AU=\frac{2}{\sqrt{1+u^2}}-\sqrt{1+u^2}=\frac{1-u^2}{\sqrt{1+u^2}}.
$$
Proceeding similarly,  $BV'=2/\sqrt{1+v^2}$ and $VV'=(1-v^2)/\sqrt{1+v^2}$. Let 
 $W$ be the point of intersection of $U'V'$ and $OC$, and set $OW=:w$. Let the extension of $U'V'$ 
  meet the extension of $AB$ at $O'$.    Menelaus's Theorem applied to $\Delta AOU$ 
 with the line $O'U'$ gives 
 $$
 \frac{w-u}{w}\cdot \frac{OO'}{OO'-1}\cdot \frac{2/\sqrt{1+u^2}}{(1-u^2)/\sqrt{1+u^2}}
 =
 \frac{UW}{OW}\cdot \frac{OO'}{AO'}\cdot \frac{AU'}{UU'}=1.
 $$
 This yields 
 \begin{equation}
  \label{eq_exercise_2015_23_10_12_00_A}
  \frac{1}{OO'}=1-\frac{2}{1-u^2}\cdot \frac{w-u}{w}.
 \end{equation}
 Similarly, Menelaus's Theorem applied to $\Delta BOV$ with the line $O'U'$ gives 
 $$
 \frac{w-v}{w}\cdot \frac{OO'}{OO'+1}\cdot \frac{2/\sqrt{1+v^2}}{(1-v^2)/\sqrt{1+v^2}}
 =
 \frac{VW}{OW}\cdot \frac{OO'}{BO'}\cdot \frac{BV'}{VV'}=1.
 $$
 This yields 
 \begin{equation}
  \label{eq_exercise_2015_23_10_12_00_B}
  \frac{1}{OO'}=\frac{2}{1-v^2}\cdot \frac{w-v}{w}-1.
 \end{equation}
 Equating the right-hand sides of \eqref{eq_exercise_2015_23_10_12_00_A} and 
 \eqref{eq_exercise_2015_23_10_12_00_B} gives 
  $\displaystyle 
 w=\frac{u+v}{1+uv}. 
 $ 

\section{A projective geometric proof}

\noindent We recall the notion of the cross ratio in projective geometry. If $A,B,C,D$ are collinear points 
that are projected along four concurrent lines meeting at $P$, to the collinear points $A',B',C',D'$, respectively, then 
we know that the cross ratio is preserved, that is, 
$$
(A,B;C,D):= \frac{AC}{AD}\Big/ \frac{BC}{BD} =\frac{A'C'}{A'D'}\Big/ \frac{B'C'}{B'D'}=:(A',B';C',D').
$$
Recall that this is an immediate consequence of the Sine Rule for triangles, using which one can see that 
\begin{eqnarray*}
\frac{AC}{AP}=\frac{\sin \angle APC}{\sin \angle PCA},&&  
\frac{AD}{AP}=\frac{\sin \angle APD}{\sin \angle PDA},\\ 
\frac{BD}{BP}=\frac{\sin \angle BPD}{\sin \angle PDB},&& 
\frac{BC}{BP}=\frac{\sin \angle BPC}{\sin \angle PCB},
\end{eqnarray*}
and so 
$$
(A,B;C,D)=\frac{\sin \angle APC}{\sin \angle APD}\Big/ \frac{\sin \angle BPC}{\sin \angle BPD}.
$$
In light of this invariance, we refer to the cross ratio of the four concurrent lines 
instead of particular collinear points on the lines.

\begin{figure}[H]
   \center
   \psfrag{A}[c][c]{${  A}$}
   \psfrag{B}[c][c]{${  B}$}
   \psfrag{C}[c][c]{${  C}$}
   \psfrag{D}[c][c]{${  D}$}
   \psfrag{a}[c][c]{${  A'}$}
   \psfrag{b}[c][c]{${  B'}$}
   \psfrag{c}[c][c]{${  C'}$}
   \psfrag{d}[c][c]{${  D'}$}
   \psfrag{P}[c][c]{${  P}$}
   \includegraphics[width=5.4cm]{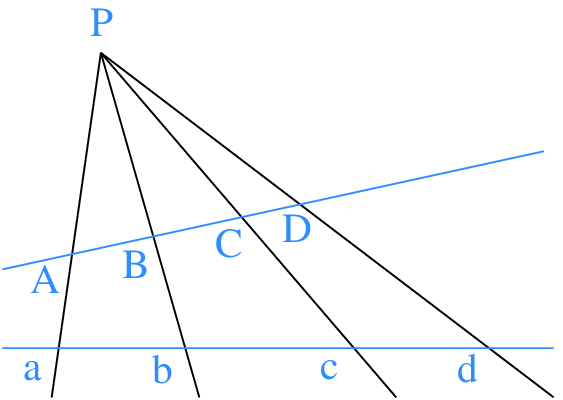}
\end{figure}

\noindent We also recall Chasles's Theorem, which says that if $A_1,A_2,A_3, A_4$ are four fixed points 
on a circle, and $P$ is a movable point, then the cross ratio of the lines 
$PA_1$, $PA_2$, $PA_3$, $PA_4$ is a constant. This is an immediate consequence of the fact that 
a chord of a circle subtends equal angles at any point on its major (or minor) arc. 

\begin{figure}[H]
   \center
   \psfrag{A}[c][c]{${  A_1}$}
   \psfrag{B}[c][c]{${  A_2}$}
   \psfrag{C}[c][c]{${  A_3}$}
   \psfrag{D}[c][c]{${  A_4}$}
   \psfrag{P}[c][c]{${  P}$}
   \includegraphics[width=4.2cm]{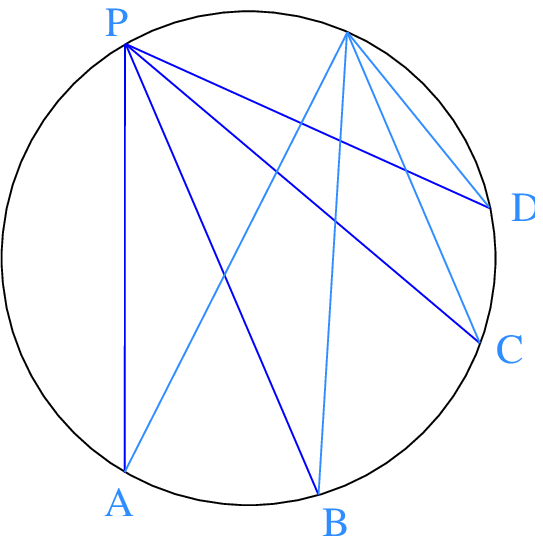}
\end{figure}

\noindent We refer to the geometric diagram for relativistic velocity addition below, 
 with the labelling of points shown. $X$ is the point of intersection of $P'A_3$ with  $OA_4$.

\begin{figure}[H]
   \center
   \psfrag{A}[c][c]{${  A_1}$}
   \psfrag{B}[c][c]{${  P}$}
   \psfrag{U}[c][c]{${ \;V}$}
   \psfrag{V}[c][c]{${  U\;}$}
   \psfrag{u}[c][c]{${  A_2}$}
   \psfrag{D}[c][c]{${  A_3}$}
   \psfrag{v}[c][c]{${  P'}$}
   \psfrag{W}[c][c]{${W}$}
   \psfrag{X}[c][c]{${  X}$}
   \psfrag{O}[c][c]{${  O}$} 
   \psfrag{C}[c][c]{${  \!\!A_4}$}
   \includegraphics[width=5.4cm]{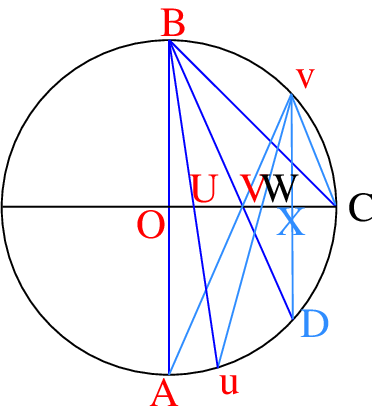}
\end{figure} 
 
 \noindent 
 As $\Delta OP'X$ is a right angled triangle, it follows that 
 $$
 OX=\cos \angle P'OX=\sin \angle POP'=\sin (2\alpha)
 =\frac{2\tan \alpha}{1+(\tan \alpha)^2} =\frac{2u}{1+u^2}.
 $$
 Hence 
 \begin{eqnarray*}
 UX&=&OX-OU=\frac{2u}{1+u^2}-u=u\cdot \frac{1-u^2}{1+u^2}
 \quad \textrm{and} \\
 WX&=&OX-OW=\frac{2u}{1+u^2}-w.
 \end{eqnarray*}
 By Chasles's Theorem, we have 
 $$
\frac{u}{1} \Big/ \frac{u-v}{1-v} =  \frac{OU}{OA_4} \Big/ \frac{VU}{VA_4} 
= 
\frac{UX}{UA_4} \Big/ \frac{WX}{WA_4}= 
\frac{u\cdot \frac{1-u^2}{1+u^2}}{1-u} \Big/ \frac{u\cdot \frac{2u}{1+u^2}-w}{1-w} .
 $$
 Solving for $w$, this yields 
  $\displaystyle 
 w=\frac{u+v}{1+uv}.
 $ 
 
 \section{A few remarks}
  
\noindent We remark that that the projective perspective also sheds light on the (algebraically easily verified) 
formula 
$$
u\oplus v=\frac{1}{u}\oplus \frac{1}{v}.
$$
Indeed, let us see the picture below, where $U', V'$ are the images of the points $U,V$, respectively, 
under  inversion in the circle.

\begin{figure}[H]
   \center
   \psfrag{A}[c][c]{${  A}$}
   \psfrag{B}[c][c]{${  B\;\;}$}
   \psfrag{U}[c][c]{${ \;V}$}
   \psfrag{V}[c][c]{${  U\;}$}
   \psfrag{u}[c][c]{${  U'}$}
   \psfrag{v}[c][c]{${  \; V'}$}
   \psfrag{P}[c][c]{${  P}$}
   \psfrag{Q}[c][c]{${Q}$}
   \psfrag{R}[c][c]{${  R}$}
   \psfrag{S}[c][c]{${  S}$} 
   \psfrag{O}[c][c]{${  O}$} 
   \includegraphics[width=7.5cm]{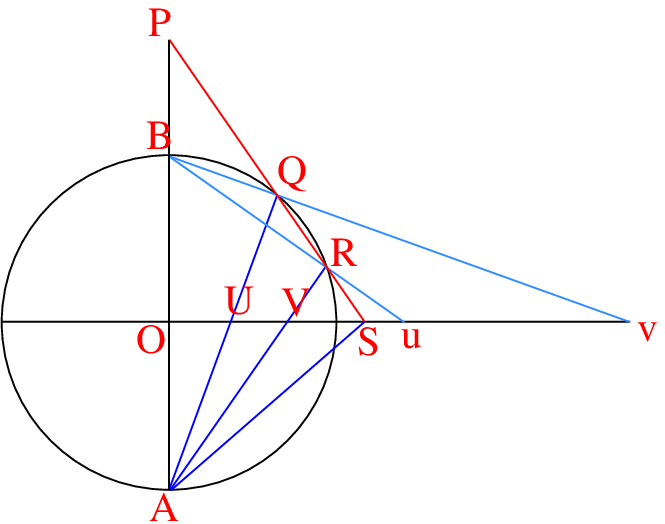}
\end{figure} 

\noindent Let $OS=:1/w$. By the preservation of the cross-ratio for  
the four collinear lines $AP, AQ, AR, AS$, we obtain 
$$
(P,Q;R,S)=(O,V;U,S)=\frac{u}{1/w}\Big/\frac{u-v}{(1/w)-v}.
$$
On the other hand, by the preservation of the cross-ratio for  
the four collinear lines $BP, BQ, BR, BS$, we obtain 
$$
(P,Q;R,S)=(O,V';U',S)=\frac{1/u}{1/w}\Big/\frac{(1/v)-(1/u)}{(1/v)-(1/w)}.
$$
Thus 
$$
\frac{u}{1/w}\Big/\frac{u-v}{(1/w)-v}=(P,Q;R,S)=\frac{1/u}{1/w}\Big/\frac{(1/v)-(1/u)}{(1/v)-(1/w)},
$$
which gives $w=\displaystyle \frac{u+v}{1+uv}$. 

We also mention that although we have been considering $u,v\in [0,1]$ for our pictures, 
one may in fact take $u,v\in [-1,1]$ without any essential change in our derivations. 
The operation $\oplus $ is associative and the set $(-1,1)$ is a group with the operation $\oplus$.


\begin{thebibliography}{99}

\bibitem{K}
Jerzy Kocik.
Geometric diagram for relativistic addition of velocities. 
{\em American Journal of Physics}, volume 80, number 8, page 737, 2012. 

  
\end{thebibliography}
\end{document}